\documentclass[conference]{IEEEtran}
\usepackage[noadjust]{cite}
\usepackage{graphicx,color}
\usepackage[dvipsnames]{xcolor}
\usepackage{amsmath,amsbsy,amsfonts,amssymb,amsthm}
\usepackage{mathrsfs,bm,bbm}
\usepackage{mathtools}
\usepackage{flushend}
\usepackage{lipsum}

\mathtoolsset{showonlyrefs}
\ifCLASSOPTIONcompsoc
\usepackage[caption=false,font=normalsize,labelfont=sf,textfont=sf]{subfig}
\else
\usepackage[caption=false,font=footnotesize]{subfig}
\fi

\IEEEoverridecommandlockouts

\newcommand{\figr}{Fig.~}
\newcommand{\secr}{Sec.~}

\usepackage{tikz}
\usetikzlibrary{shapes}

\usepackage{acro}
\DeclareAcronym{AWGN}{short = AWGN ,long = additive white gaussian noise}
\DeclareAcronym{AoI}{short = AoI ,long = age of information}
\DeclareAcronym{AoII}{short = AoII ,long = age of incorrect information}

\DeclareAcronym{CDF}{short = CDF ,long = cumulative distribution function}
\DeclareAcronym{CRA}{short = CRA ,long = contention resolution ALOHA}
\DeclareAcronym{CRDSA}{short = CRDSA ,long = contention resolution diversity slotted ALOHA}
\DeclareAcronym{CSA}{short = CSA ,long = coded slotted ALOHA}
\DeclareAcronym{C-RAN}{short = C-RAN ,long = cloud radio access network}
\DeclareAcronym{DAMA}{short = DAMA ,long = demand assigned multiple access}
\DeclareAcronym{DSA}{short = DSA ,long = diversity slotted ALOHA}
\DeclareAcronym{eMBB}{short = eMBB ,long = enhanced mobile broadband}
\DeclareAcronym{FEC}{short = FEC ,long = forward error correction}
\DeclareAcronym{GEO}{short = GEO ,long = geostationary orbit}
\DeclareAcronym{GF}{short = GF ,long = generating function}
\DeclareAcronym{IC}{short = IC ,long = interference cancellation}
\DeclareAcronym{IoT}{short = IoT ,long = Internet of things}
\DeclareAcronym{IRSA}{short = IRSA ,long = irregular repetition slotted ALOHA}
\DeclareAcronym{KPI}{short = KPI ,long = key performance indicator}
\DeclareAcronym{LEO}{short = LEO ,long = low Earth orbit}
\DeclareAcronym{MAC}{short = MAC ,long = medium access}
\DeclareAcronym{mMTC}{short = mMTC ,long = massive machine-type communications}
\DeclareAcronym{MC}{short = MC ,long = Markov chain}
\DeclareAcronym{PDF}{short = PDF ,long = probability density function}
\DeclareAcronym{PER}{short = PER ,long = packet error rate}
\DeclareAcronym{PLR}{short = PLR ,long = packet loss rate}
\DeclareAcronym{PMF}{short = PMF ,long = probability mass function}
\DeclareAcronym{RA}{short = RA ,long = random access}
\DeclareAcronym{rv}{short = r.v. ,long = random variable}
\DeclareAcronym{SA}{short = SA , long = slotted ALOHA}
\DeclareAcronym{SIC}{short = SIC ,long = successive interference cancellation}
\DeclareAcronym{SNR}{short = SNR ,long = signal-to-noise ratio}
\DeclareAcronym{SFG}{short = SFG ,long = signal flow graph}
\DeclareAcronym{TDM}{short = TDM ,long = time division multiplexing}
\begin{document}

\title{\huge Spatio-Temporal Information Freshness for\\ Remote Source Monitoring in IoT Systems}
\author{
\IEEEauthorblockN{Andrea Munari, Federico Chiariotti, Leonardo Badia, Petar Popovski
\thanks{A. Munari (andrea.munari@dlr.de) is with the Inst. of Comm. and Navigation, German Aerospace Center (DLR), Germany. F. Chiariotti (federico.chiariotti@unipd.it) and L. Badia (leonardo.badia@unipd.it) are with the Dept. of Information Eng., Univ. of Padova (Italy), P. Popovski (petarp@es.aau.dk) is with the Dept. of Electronic Systems, Aalborg University (Denmark). This work was supported in part by the Velux Foundation, Denmark, through the Villum Investigator Grant WATER, nr. 37793. A. Munari acknowledges the financial support by the Federal Ministry of Research, Technology and Space of Germany in the programme of “Souverän. Digital. Vernetzt.” Joint project 6G-RIC, project id. number: 16KISK022.}
}
}
\date{}
\maketitle
\thispagestyle{empty}
\pagestyle{empty}

\begin{abstract}
    The widespread adoption of \ac{AoI} as a meaningful and analytically tractable information freshness metric has led to a wide body of work on the timing performance of \ac{IoT} systems. However, the spatial correlation inherent to environmental monitoring has been mostly neglected in the recent literature, due to the significant modeling complexity it introduces. In this work, we address this gap by presenting a model of spatio-temporal information freshness, considering the conditional entropy of the system state in a remote monitoring scenario, such as a low-orbit satellite collecting information from a wide geographical area. Our analytical results show that purely age-oriented schemes tend to select an overly broad communication range, leading to inaccurate estimates and energy inefficiency, both of which can be mitigated by adopting a spatio-temporal approach.
\end{abstract}

\newtheorem{prop}{Proposition}
\newtheorem{lemma}{Lemma}
\newtheorem{remark}{Remark}

% maths and probability
\newcommand{\pr}{\ensuremath{\mathsf P}}
\newcommand{\expOp}{\ensuremath{\mathbb E}}
\newcommand{\de}{\mathrm{d}}
\newcommand{\given}{\, | \,}
\newcommand{\givenS}{\ensuremath{\vert}}
\newcommand{\norm}[2]{\left \lVert #1 \right \rVert_{#2}}
% PPP
\newcommand{\pppDens}{\ensuremath{\rho}}
\newcommand{\area}{\ensuremath{\mathcal A}}
\newcommand{\rad}{\ensuremath{\mathsf R}}
\newcommand{\radMax}{\ensuremath{\rad_{\mathsf m}}}
\newcommand{\powLaw}{\ensuremath{\alpha}}

% Markov chains
\newcommand{\pTrans}{\ensuremath{q}}

\newcommand{\Mc}{\ensuremath{X}}
\newcommand{\Mcn}{\ensuremath{X_n}}
\newcommand{\Rc}{\ensuremath{Y}}
\newcommand{\rc}{\ensuremath{y}}
\newcommand{\Rcn}{\ensuremath{\Rc_n}}
\newcommand{\rcn}{\ensuremath{\rc_n}}
\newcommand{\Distn}{\ensuremath{D_n}}
\newcommand{\distn}{\ensuremath{d_n}}
\newcommand{\Est}{\ensuremath{\hat{X}}}
\newcommand{\Estn}{\ensuremath{\hat{X}_n}}
\newcommand{\mc}{\ensuremath{x}}
\newcommand{\mcn}{\ensuremath{\mc_n}}
\newcommand{\est}{\ensuremath{\hat{x}}}
\newcommand{\estn}{\ensuremath{\hat{x}_n}}

% channel access
\newcommand{\pTx}{\ensuremath{\zeta}}
\newcommand{\ps}{\ensuremath{\mathsf{p_s}}}
\newcommand{\peras}{\ensuremath{\varepsilon}}

% transition and stationary probabilities
\newcommand{\qZO}{\ensuremath{\alpha}}
\newcommand{\qOZ}{\ensuremath{\beta}}
\newcommand{\asymm}{\ensuremath{\eta}}
\newcommand{\statZ}{\ensuremath{\pi_0}}
\newcommand{\statO}{\ensuremath{\pi_1}}
\newcommand{\statZZ}{\ensuremath{\pi_{0,0}}}
\newcommand{\statOO}{\ensuremath{\pi_{1,1}}}
\newcommand{\statZO}{\ensuremath{\pi_{0,1}}}
\newcommand{\statOZ}{\ensuremath{\pi_{1,0}}}

\newcommand{\nodes}{\ensuremath{\mathsf m}}

% basic metrics
\newcommand{\tru}{\ensuremath{\mathsf S}}
\newcommand{\load}{\ensuremath{\mathsf G}}
\newcommand{\Agen}{\ensuremath{\Delta_n}}
\newcommand{\agen}{\ensuremath{\delta_n}}
\newcommand{\ent}{\ensuremath{H}}
\newcommand{\condent}{\ensuremath{\mathsf h}}

% AoI related metrics
\newcommand{\overbar}[1]{\mkern 1.5mu\overline{\mkern-2mu#1\mkern-7mu}\mkern .5mu}
\newcommand{\overbara}[1]{\mkern 1.5mu\overline{\mkern0.1mu#1\mkern-0.1mu}\mkern 1.5mu}

\section{Introduction} \label{sec:intro}

Over the past decade, billions of \ac{IoT} sensors have been deployed to enable a plethora of different applications, from industrial automation to environmental monitoring~\cite{ericsson2024mobility}. The unique challenges of dealing with massive numbers of low-power, low-throughput devices have reshaped medium access design~\cite{guo2021enabling}, as well as the performance metrics, which fundamentally differ from those used for human communication.

Rather than aiming at minimizing the latency of every single packet, \ac{AoI}~\cite{Kaul11_SECON} has emerged as a compact and useful metric for \ac{IoT} scenarios. The metric measures the freshness of the last update that was correctly received, offering a better proxy for the alignment between the estimate available to the receiver and the real state of the environment. 
Since its introduction, \ac{AoI} and its optimization have been the subject of a vast body of scientific literature~\cite{Yates20_Survey}, as well as several extensions.

While the timing aspects of information freshness have been thoroughly investigated, the spatial dimension is so far mostly unexplored, or dealt with only as part of the medium access model~\cite{yang2021spatiotemporal}: research that considers both dimensions, exploiting both age-related notions and spatial correlations in the physical process measured by the sensors, is still relatively sparse. The first work to explicitly model correlated sources~\cite{kalor2022timely} used beliefs at the receiver to compute a spatio-temporal \ac{AoI} metric and optimize receiver-driven scheduling of sensor observations. Another model, which starts from similar considerations, was presented in~\cite{zancanaro2023modeling}: in this case, simpler access strategies are considered, which allow for a more complete freshness analysis.

Other possible models adapt the definition of \ac{AoI} without explicitly considering the receiver's beliefs: it is possible to reset the age to a value that reflects the correlation of the received measurement with the intended target, artificially increasing it to account for imperfect measurements, or to adapt the age function~\cite{tong2022age}, considering a distance-based weighting function so that age increases faster for less reliable measurements~\cite{fidler20242d}. Sensor correlation is also a critical factor when considering more complex environment models such as digital twins~\cite{bellavista2024odte}.

\begin{figure}[t]
    \centering   
    \tikzset{pics/.cd,
  sensor/.style={
    code={
        \draw[pic actions] (-0.04,-0.115) rectangle (0.04,0.025);
        \draw (0,0.025) -- (0,0.085);
        \draw (0,0.085) -- (0.03,0.115);
        \draw (0,0.085) -- (-0.03,0.115);
    } 
  },
  tree/.style={
    code={
        \draw[fill=brown] (-0.03,-0.1) rectangle (0.03,-0.3);
        \node [cloud, inner sep=2pt,draw,minimum width = 0.25cm,fill=green!60, aspect=2] {};
    } 
  },
  satellite/.style={
    code={
        \draw (0.07,-0.07) -- (0.17,-0.17);
        \draw[pic actions,rotate around={45:(0.27,0.27)}] (0.095,0.17) rectangle (0.445,0.37);
        \draw (0.38,0.24) -- (0.24,0.38);
        \draw (0.3,0.16) -- (0.16,0.3);
        \draw (0.07,0.07) -- (0.39,0.39);
        \draw[pic actions,rotate around={45:(-0.27,-0.27)}] (-0.095,-0.17) rectangle (-0.445,-0.37);
        \draw (-0.38,-0.24) -- (-0.24,-0.38);
        \draw (-0.3,-0.16) -- (-0.16,-0.3);
        \draw (-0.07,-0.07) -- (-0.39,-0.39);        
        \draw[pic actions] circle[radius=1mm];
        \draw (0,0) -- (-0.17,0.17);
    } 
  }
}

\begin{tikzpicture}

% \draw (0,0)[draw=none,fill=red!0] ellipse (2cm and 1cm);
\draw (0,0)[black,fill=blue!15] ellipse (1.5cm and 0.75cm);
\draw (0,0)[black,dashed,fill=blue!30] ellipse (1cm and 0.5cm);
\draw (0,0)[black,dashed,fill=blue!45] ellipse (0.5cm and 0.25cm);
\draw (1.38,0.3) -- (0,2);
\draw (-1.38,0.3) -- (0,2);

\draw (0,0.2) pic[draw,fill=green!20]{tree};
\draw (-0.25,2.25) pic[draw,fill=gray!20]{satellite};

\draw (1.2,0) pic[draw,fill=gray!20]{sensor};
\draw (0.6,-0.2) pic[draw,fill=gray!20]{sensor};
\draw (0.4,0.3) pic[draw,fill=gray!20]{sensor};
\draw (0.6,0.85) pic[draw,fill=gray!20]{sensor};
\draw (-0.1,-0.9) pic[draw,fill=gray!20]{sensor};
\draw (-1.1,-0.15) pic[draw,fill=gray!20]{sensor};
\draw (-0.4,-0.58) pic[draw,fill=gray!20]{sensor};
\draw (1.75,0.15) pic[draw,fill=gray!20]{sensor};
\draw (-1.55,-0.45) pic[draw,fill=gray!20]{sensor};
\draw (-1.25,0.65) pic[draw,fill=gray!20]{sensor};
\draw (-0.25,0.05) pic[draw,fill=gray!20]{sensor};

\end{tikzpicture}
    \caption{A schematic of the considered application scenario.}
    \vspace{-1em}
    \label{fig:schematic}
\end{figure}
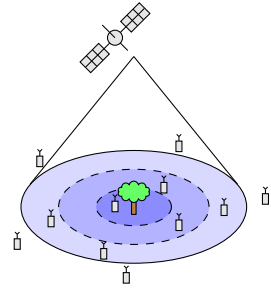

In this work, we consider an application scenario such as the one depicted in Fig.~\ref{fig:schematic}: the physical process that the network aims at measuring, represented by the tree, is localized in space, and only sensors placed very close to it may measure it exactly. However, we cannot directly control sensor placement, and energy efficiency considerations prevent us from only considering very close sensors, as it would quickly deplete their batteries. The correlation of each sensor's observations with the real state of the process depends on the distance between them, represented by the shaded areas in the figure. Measurements from sensors closer to the tree are highly correlated to its state, while sensors that are placed too far from it can only make very unreliable observations. The receiver, represented by the satellite, can collect observation within a controllable radius and track the process. In this case, pure \ac{AoI} is insufficient to represent the freshness of the information available to the satellite: similarly to the model from~\cite{kalor2022timely}, we need to consider both the freshness of updates and their uncertainty to account for unreliable measurements.

Unlike~\cite{kalor2022timely}, which studies a centralized scheduling scheme in which the receiver perfectly knows the placement of the sensors, we consider a slotted ALOHA access protocol, which is more realistic for \ac{IoT} systems, and assume that nodes do not have any geolocation information. In this case, the uncertainty over the state of the source process can be characterized by the conditional entropy at the receiver. In this work, we consider a \emph{forgetful} receiver that only tracks the age and the last received value. This makes the analysis tractable, as well as allowing for implementation on constrained hardware. We leave more complex setups for future work.

We provide a novel closed-form analysis of the conditional entropy, which reveals the importance of including a spatial component in information freshness problems: the minimum entropy solution is obtained by setting a much narrower radius than the \ac{AoI}-optimal one, reducing energy consumption by limiting the number of polled sensors and improving the accuracy of the remote estimation. The optimization of the collection radius also has significant practical implications, offering hints on system dimensioning. 

\emph{Notation:} We denote vectors and matrices in lower- and upper-case boldface, respectively, e.g., $\bm a$, $\mathbf A$. In particular, a vector denoting the position of a point on $\mathbb R^2$ is indicated as $\bm x$. A discrete random variable (r.v.) and its realization use upper- and lower-case, respectively, e.g., $X$ and $x$. Its probability mass function is denoted as $p_X(x){=} \mathsf P(X{=}x)$, and the subscript is omitted for brevity whenever it can be inferred from the context. The notation is extended straightforwardly to conditional distributions. We express the indicator function as $\mathbbm 1(x,y)$, taking value $1$ if $x=y$ and $0$ otherwise, and we denote the complementary of a binary value $y{\in}\{0,1\}$ as $\overbara{y}$.

\section{System Model}
\label{sec:sysModel}

Throughout our study, we focus on a set of IoT nodes (devices), uniformly distributed over the circular coverage area $\area\subseteq \mathbb R^2$ of a receiver. Without loss of generality, \area\ is centered at the origin of the plane, and has radius \radMax. For a density \pppDens\ [nodes/unit area], we consider $\nodes = \pppDens\pi  \radMax^2$ devices to populate the system, observing a physical process of interest and reporting updates to the receiver. The process is modeled as a discrete-time, two-state Markov chain \Mcn, taking values in $\mathcal X = \{0,1\}$. The one-step transition matrix for the chain is 
\begin{align}
    \mathbf A = 
    \left(
        \begin{array}{cc}
            1-\pTrans   & \pTrans \\  
            \eta\pTrans & 1-\eta\pTrans
        \end{array}
    \right)
\end{align}
where $\eta>0$ is referred to as \emph{asymmetry factor}. 

Devices share a common wireless channel towards the receiver by means of a slotted ALOHA access policy \cite{Abramson77:PacketBroadcasting}. Specifically, time is divided in slots of equal duration, corresponding to the time steps of the Markov process, and at each slot $n\in\mathbb N$: i) \Mcn\ transitions following $\mathbf A$, and ii) every device independently decides with probability \pTx\ whether to observe the process and transmit the obtained value. A sent packet is erased with probability \peras, bringing no power contribution to the receiver, or arrives unfaded with probability $1-\peras$.\footnote{The considered model captures setups in which packets coming from different nodes are equally likely to be retrieved. For instance, this is the case under power control, or in settings where the relative displacement of nodes is not predominant on the overall path-loss, e.g., for a LEO satellite.}  Resorting to the well-established collision channel model, we assume that a slot where two or more unerased messages reach the receiver does not allow retrieval of information, whereas a singleton slot leads to successful decoding \cite{Abramson77:PacketBroadcasting}.Accordingly, each slot sees delivery of a message with probability
\begin{align}
\ps = \nodes\,\pTx (1-\peras) \cdot[1-\pTx(1-\peras)]^{\nodes-1}.
\label{eq:ps}
\end{align}

Due to the displacement of devices within \area, the reliability of the observations produced by a node depends on its position. In particular, we assume that devices closer to the origin of the plane provide more precise readings, whereas the accuracy of messages coming from farther away progressively reduces, as the longer the information about the event propagates physically, the higher the probability that it may be distorted. Such a setting captures, for instance, the case of an application tracking the value of a physical quantity at the specific coordinates of interest (plane origin), or of a monitoring system that has to detect an event (e.g., presence/absence of an object) in the corresponding surroundings. To model this, we partition the coverage area \area\ in $K$ regions: % 
\begin{align}
    \area_i := \{ \bm x \in \area: i \rad \leq \norm{\bm x}{2} < (i+1)\rad  \,\}
\end{align}
for $i\in \mathcal K = \{0,\dots,K-1\}$, and where $\rad:=\radMax/K$. Whenever a device in $\area_i$ transmits an update, its message contains the correct value of \Mcn\ with probability $\lambda(i)$, or the erroneous reading $\overbar{\Mcn}\,$ with probability $1-\lambda(i)$. In the remainder of our discussion we set 
\begin{align}
    \lambda(i) = (1+i\rad)^{-\powLaw}
    \label{eq:powerLaw}
\end{align}
 with $\alpha > 0$, although the analysis we present can be adapted straightforwardly to different reliability functions.
 
 At the receiver side, we assume that no knowledge can be gathered on the area out of which an incoming message was generated, e.g., due to location-unaware devices, or to the fact that no geographical information is added to the packet payload. Moreover, we consider the receiver to be \emph{forgetful}, i.e., that it may estimate the state of the process of interest solely based on the last obtained reading. This solution is inspired by practically relevant IoT setups, epitomizing the common behavior of an application layer which is fed with the payload of incoming messages, and does not lean on any other cross-layer information exchange or memory. More intelligent models, which also require a higher computational power, will be considered as future work.
 Following this approach, the receiver at time $n$ has knowledge about: \textit{(i)} the time elapsed since the last reception, i.e., the age of information \Agen\ \cite{Yates20_Survey}; and \textit{(ii)} the content of the last retrieved message, which we denote by \Rcn. Without loss of generality, the reception of a message resets the AoI value to $\Agen=0$.

\begin{figure}
    \hspace{-1em}
    \subfloat[coverage range $\radMax = 6\rad$]{
        \includegraphics[width=.48\columnwidth]{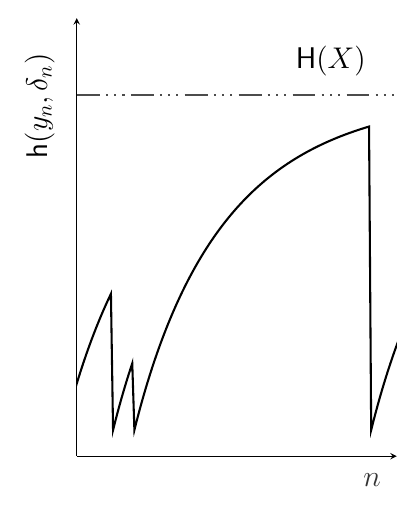}}        
    \hspace*{.1em}
    \subfloat[coverage range $\radMax = 15\rad$]{
        \includegraphics[width=.48\columnwidth]{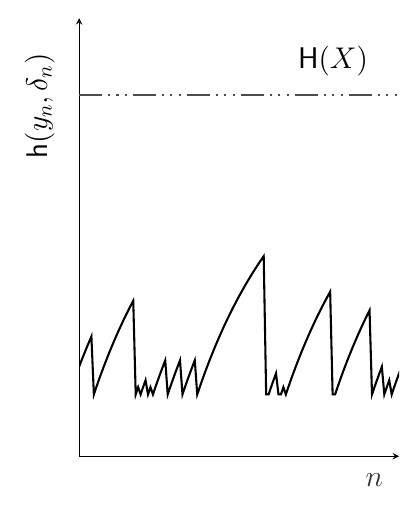}}
    \caption{Examples of time evolution of $\condent(\rcn,\agen)$ when tracking a symmetric source. The horizontal line reports the value of the stationary source entropy, $\mathsf H(X) = -\pi_0\log_2 \pi_0 - \pi_1 \log_2 \pi_1$.
    The plot was generated for $\alpha{=}0.02$, $\rad{=}10$, $q{=}0.005$, $\rho{=}0.05$, $\pTx{=}5{\cdot}10^{-4}$, $\peras{=}0.1$.}
    \label{fig:timeline}
    \vspace{-1em}
\end{figure}

 The uncertainty on the current state of the process \Mcn, conditioned on the available information, is then naturally captured by the conditional entropy
\begin{align}
    \condent(\rcn,\agen) := H(\Mcn\given \Rcn=\rcn,\,\Agen=\agen). 
    \label{eq:h}
\end{align}
The quantity is relevant from an operational standpoint, as it may serve as basis to properly gauge the current knowledge or make actuation decisions, and is akin to the age of uncertainty introduced in \cite{Liew22_TIT}. Two examples of how $\condent(\rcn,\agen)$ may evolve over time are reported in \figr\ref{fig:timeline}. In the first case, (a), a smaller radius of $\radMax=6\rad$ was considered, whereas the rightmost plot, (b), was obtained for a wider coverage of $\radMax=15\rad$. The figure hints at a key tradeoff. On the one hand, selecting a smaller radius leads to more sporadic update deliveries (corresponding to resets of $\condent(\rcn,\agen)$), and the uncertainty tends to grow for longer periods of time.

Whenever the receiver does not retrieve messages for some time (central part of plot (a)), its uncertainty tends to converge to the stationary entropy of the source, i.e., \mbox{$\mathsf H(X) = -\pi_0 \log_2 \pi_0 - \pi_1 \log_2 \pi_1$}. This effect is much more mitigated in figure (b), as the presence of more nodes in the system results in an improvement of the frequency of message delivery, with a lower AoI and an uncertainty that is more often brought back to its minimum. On the other hand, such reset value gets higher as \radMax\ increases. Indeed, incoming messages are more likely to be generated by devices which are farther away from the origin, containing less reliable readings as per \eqref{eq:powerLaw}, and leaving the receiver with a higher residual uncertainty. The critical balance of these two factors will determine system performance, and its analytical characterization represents the core contribution of this paper. We will provide in \secr\ref{sec:analysis} a closed-form derivation of the conditional entropy
\begin{align}
    H(\Mcn\given \Rcn,\Agen) =\!\! \sum\nolimits_{\rcn,\agen} \!\!\condent(\rcn,\agen) \,p(\rcn,\agen)
    \label{eq:condEntropy}
\end{align}
as well as the cumulative distribution function of $\condent(\rcn,\agen)$.

\section{Analysis}
\label{sec:analysis}

Let us start by considering a slot $n$ in which the receiver successfully retrieves a message, i.e., the \ac{AoI} is reset to {$\Agen{=}0$}.
Applying Bayes' rule, the distribution of the state of the tracked process conditioned on receiving a reading with value \Rcn\ can be expressed as 
\begin{align}
    \!\!\!\!\!\!\!p_{\Mcn\givenS \Rcn,\Agen}(\mcn\given \rcn,0) \!=\! \frac{p_{\Rcn\givenS\Mcn,\Agen}(\rcn\given \mcn,0) \, \pi_{\mcn}}{\sum\limits_{\mcn^{\prime}\in \mathcal X} p_{\Rcn\givenS\Mcn,\Agen}(\rcn\given \mcn^{\prime},0) \, \pi_{\mcn^{\prime}}}
    \label{eq:pXnGivenYnReset}
\end{align}
where $\pi_{\mcn}$ is the stationary distribution of \Mcn, readily obtained as $\pi_0 = \eta/(1+\eta)$, $\pi_1 = 1/(1+\eta)$, and we have used the fact that event of successfully receiving a packet is independent of \Mcn\ for the transmission policies under study.

In turn, $p(\rcn\given \mcn,0)$ can be conveniently expressed in terms of the r.v. $\Distn \in\mathcal K$, capturing the region the message originated from. Leaning on the law of total probability, we have
\begin{align}
    p(\rcn\given \mcn,0) = \sum_{\distn=0}^{K-1} p(\rcn\given \mcn,0,\distn) \, p(\distn\given \mcn,0).
    \label{eq:pYnGivenXnReset}
\end{align}
Within \eqref{eq:pYnGivenXnReset}, the term $p(\distn\given\mcn,0)$ accounts for the probability of a received message to be generated from region \distn. 
Observing that the devices are uniformly distributed within \area, and transmit independently of their location as well as of the state of the process, the quantity can be computed as
\begin{align}
    \!\!\!\!\!p(\distn\given \mcn,0) = \!\int_{\distn\rad}^{(\distn+1)\rad} \frac{2r}{\radMax^2} \, dr = \frac{2\distn+1}{K^2}.
    \label{eq:pDnGivenXn}
\end{align}
The first factor within the summation in \eqref{eq:pYnGivenXnReset} captures the probability of a decoded message containing value \rcn, conditioned on the actual state of the process and on the reliability of the sender. Accordingly, we can write in a compact form
\begin{align}
    p(\rcn\given\mcn,0,\distn) =  \lambda(\distn) \,\mathbbm{1}(\mcn,\rcn) 
     + (1{-}\lambda(\distn))\, \mathbbm{1}(\bar{\mc}_n,\rcn).
     \label{eq:pYnGivenDnXn}
\end{align}
For the power-law in \eqref{eq:powerLaw}, combining \eqref{eq:pYnGivenDnXn} and \eqref{eq:pDnGivenXn} we obtain
\begin{align}
    p(\rcn\given \mcn,0) = \sum_{\distn=0}^{K-1} \frac{2\distn+1}{K^2(1+\distn \, \rad)^{\alpha}}, \quad \text{if } \rcn=\mcn
    \label{eq:pYnGivenXnReset_law}
\end{align}
and its complementary value for the case $\rcn\neq \mcn$.
Plugging \eqref{eq:pYnGivenXnReset_law} into \eqref{eq:pXnGivenYnReset} thus provides the statistics available at the receiver on the state of the process of interest upon reception of a packet. The result, in turn, allows us to derive the general probability $p(\mcn\given \rcn,\agen)$ for any $\agen>0$, obtained as the $\agen$-step evolution of the transition matrix $\mathbf A$ with initial distribution $p_{\Mcn\givenS\Rcn,\Agen}(0\given\rcn,0)$ and $p_{\Mcn\givenS\Rcn,\Agen}(1\given\rcn,0)$.

The uncertainty that the forgetful receiver has on the current state of the process given the currently available knowledge can finally be computed by definition as
\begin{align}
    \condent(\rcn,\agen) = -\!\!\sum_{\mcn\in\mathcal X} p(\mcn\given\rcn,\agen) \log_2 p(\mcn\given\rcn,\agen).
    \label{eq:condEntropy_def}
\end{align}
To conclude, we also derive the conditional entropy $H(\Mcn\given\Rcn,\Agen)$. From \eqref{eq:condEntropy}, this requires the PMF $p(\rcn,\agen)$, capturing, for a generic time slot $n$, the joint probability of having decoded the last message \agen\ slots ago, and of the received reading being \rcn. Observing that packet receptions are i.i.d. across slots with probability \ps, the current \ac{AoI} value is independent of the content of the incoming message, as each node's transmission probability does not depend on its age or position.
Accordingly, $p(\rcn,\agen) = p(\rcn) p(\agen)$. Moreover, $p(\rcn)$ can simply be computed from \eqref{eq:pYnGivenXnReset} as
\begin{align}
    p(\rcn) = \sum\nolimits_{\mcn\in\mathcal X} p(\rcn\given \mcn,0) \pi_{\mcn}.
    \label{eq:pYnStat}
\end{align}
In turn, the receiver experiences an \ac{AoI} $\agen\geq 0$ at the start of slot $n$ if the last success was followed by $\agen-1$ failures:
\begin{align}
    p(\agen) = \ps (1-\ps)^{\agen-1}.
    \label{eq:pAgenStat}
\end{align}
Combining \eqref{eq:pYnStat}-\eqref{eq:pAgenStat} with \eqref{eq:condEntropy_def} provides the conditional entropy.

\emph{Remark:} The approach followed in the analysis above considers an i.i.d. distribution for the random variable $D_n$ across slots. This implicitly corresponds to independently re-drawing the positions of all nodes at each time unit, and thus introduces an approximation of the behavior of a practical topology in which the locations of the devices are static. The validity of this assumption will be discussed in Sec.~\ref{sec:results}. 
\section{Results and Discussion}
\label{sec:results}

To gauge the impact of spatio-temporal information freshness on the forgetful receiver, we focus on a population of devices spread over \area\ with density $\rho = 5\cdot 10^{-2}$ [nodes/m$^2$], transmitting at every slot with probability $\pTx = 10^{-4}$. A sent packet is erased with probability $\peras = 0.1$, and the tracked Markov process \Mcn\ transitions from state $0$ to $1$ with probability $q=5\cdot 10^{-3}$. Unless otherwise specified, we assume that nodes produce readings following the reliability law in \eqref{eq:powerLaw}, where $\alpha = 0.02$, and $\rad=10$m. 

\begin{figure}
    \centering
    \includegraphics[width=.9\columnwidth]{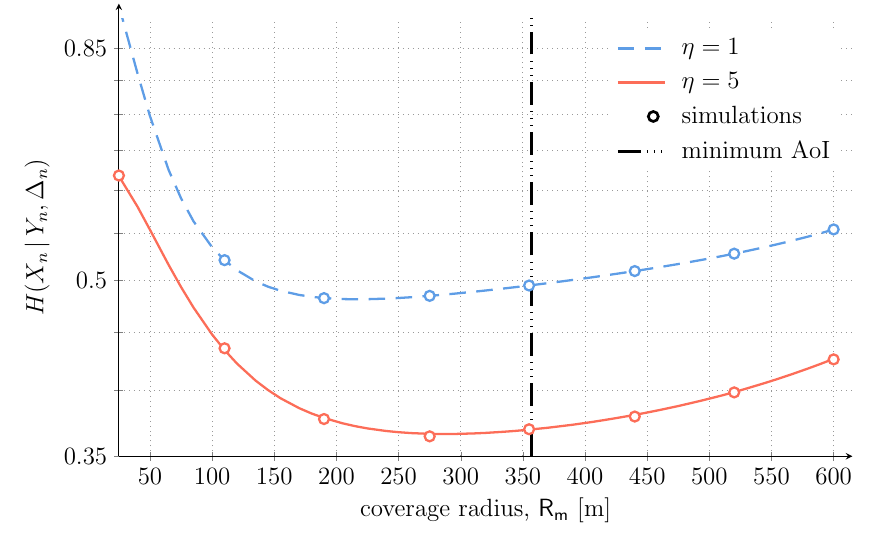}
    \caption{Conditional entropy $H(\Mcn\given \Rcn,\Agen)$ vs. coverage radius $\radMax$, when tracking a symmetric (dashed line) or an asymmetric (solid line) process. The markers report the outcomes of simulation results. The vertical line indicates the coverage radius that would minimize \ac{AoI}.}
    \label{fig:HvsRad}
\end{figure}

We start our discussion by reporting in \figr\ref{fig:HvsRad} the conditional entropy $H(\Mcn\given\Rcn,\Agen)$ against the coverage radius \radMax. The dashed line reports the case of monitoring of a symmetric process ($\eta=1$), whereas the solid curve was obtained for an asymmetric case with $\eta = 5$. Circle markers show the results of detailed network simulations, aimed at verifying the approximation introduced in our analysis (see Remark $1$). Specifically, for each simulation, the positions of the $\nodes$ nodes were uniformly drawn within \area, and their location was kept fixed for the whole duration of the run ($10^6$ slots). Throughout the run, the Markov process evolved according to $\mathbf A$, and the time evolution of $\condent(\rcn,\agen)$ was computed via \eqref{eq:condEntropy_def} to derive its average value. Each of the reported marked points shows in turn the average of $20$ independent topologies. The excellent match between analysis and simulations under all conditions confirms the tightness of the modeling assumptions. 

More interestingly, \figr\ref{fig:HvsRad} reveals the existence of an optimal coverage range to minimize $H(\Mcn\given\Rcn,\Agen)$. When coverage is too small, updates are sporadic as few nodes fall within \area. In this case, even if incoming messages are highly reliable in view of the proximity of active devices to the origin, the receiver's uncertainty is dominated by the lack of timely (fresh) updates. On the other hand, if the coverage radius is increased beyond a critical point $\radMax^*$, the conditional entropy grows once more (rightmost region of the plot). In such conditions, even if messages are received more often, their reliability decreases sharply due to the higher probability of obtaining readings from nodes farther away from the origin, leading to a higher uncertainty. Striking a proper balance between the temporal and spatial components of information freshness is thus paramount, and shall represent a key criterion in the design of the system. From this standpoint, it is relevant to remark that the optimal radius $\radMax^*$ can be significantly smaller that the one that would be obtained considering \ac{AoI} only. This is reported for reference by the vertical dash-dotted line in the figure. For the generate-at-will case under study, such configuration is obtained by operating the system at maximum throughput \cite{Yates17:AoI_SA,Munari21_TCOM_AoI}, i.e., a channel load of $1$ [pkt/slot], with a corresponding radius $(\pi\! \rho \pTx \peras)^{-1/2}$.

\figr\ref{fig:HvsRad} also highlights a difference in the optimal coverage when tracking symmetric and asymmetric sources, with the latter case presenting lower conditional entropy and a larger $\radMax^*$. As $\eta$ increases, the tracked process spends more time in state $0$, and the uncertainty on its current conditions tends to converge more quickly to a lower value even in the absence of incoming updates. This effect reduces the impact of the spatial component of information freshness, as a low level of reliability of incoming messages can be compensated by the the knowledge of the biased behavior of \Mcn.

\begin{figure}
    \subfloat[]{
        \includegraphics[width=.5\columnwidth]{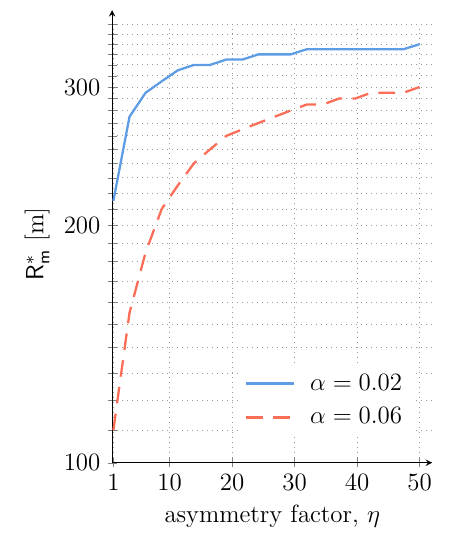}}        
    \hspace*{.1em}
    \subfloat[]{
        \includegraphics[width=.48\columnwidth]{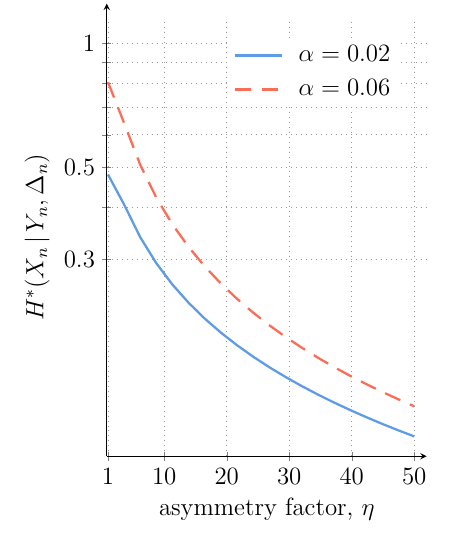}}
    \caption{(a) coverage range $\radMax^*$ minimizing the conditional entropy of the forgetful receiver; (b) corresponding minimum value of $H(\Mcn\given\Rcn,\Agen)$.}
    \label{fig:optHandRadius}
    \vspace{-1em}
\end{figure}

To further delve into this aspect, we report in \figr\ref{fig:optHandRadius} the optimal coverage radius $\radMax^*$, (a), and the corresponding minimum conditional entropy $H^*(\Mcn\given\Rcn,\Agen)$, (b), against the asymmetry factor. The optimal radius sharply increases as the asymmetry grows, and eventually, $\radMax^*$ tends to converge to the value minimizing \ac{AoI}, i.e., $(\pi\! \rho \pTx \peras)^{-1/2}$, as the temporal component of freshness dominates.  A corresponding sharp decrease in the attainable average uncertainty is experienced as $\eta$ grows, obtained by admitting more nodes in the system.  The plots report trends for two different values of the exponent $\alpha$, determining the reliability of incoming messages as per \eqref{eq:powerLaw}. It is interesting to note that, for larger values of $\alpha$, the optimal coverage shrinks, as an effect of the quicker degradation with distance of the quality of the updates. In this setting, trading off messages' timeliness for an increased reliability pays off. For instance, when $\eta=5$ (same setup considered in \figr\ref{fig:HvsRad}), the optimal coverage radius reduces to less than one half of what would be optimal in terms of  when $\alpha=0.06$.

\begin{figure}
    \includegraphics[width=.9\columnwidth]{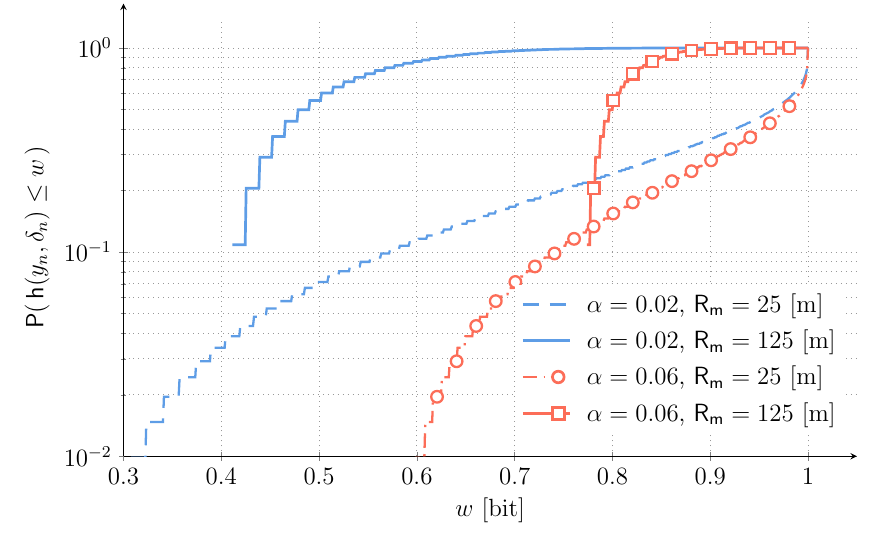}
    \caption{Cumulative distribution function of the uncertainty $\condent(\rcn,\agen)$ at the receiver. Results obtained for a symmetric process \Mcn, and for $\alpha=0.02$.}
    \label{fig:cdf}
    \vspace{-1em}
\end{figure}

We conclude our discussion by reporting in \figr\ref{fig:cdf} the cumulative distribution function of the uncertainty at the receiver at a generic point in time, conditioned on the last received message and its \ac{AoI}. The metric can be obtained from \eqref{eq:condEntropy_def}, \eqref{eq:pYnStat}-\eqref{eq:pAgenStat} as
\begin{align}
    \mathsf P\big( \condent(\rcn,\agen) \leq w \big) = \sum_{\substack{(\rcn,\agen) \text{ s.t.}\\ \condent(\rcn,\agen)\leq w}} p(\rcn,\agen).
\end{align}
In the plot, we focus on a symmetric process \Mcn, and consider two different coverage radiuses: a relatively small one ($25$ [m], dashed lines) and a larger one ($125$ [m], solid lines). Curves without markers were obtained for $\alpha=0.02$, whereas the ones with markers denote the behavior with $\alpha=0.06$. Consider first the former case. When a narrower \area\ is covered, the receiver's uncertainty is reset to a smaller value whenever a message is successfully decoded, by virtue of the more reliable readings being collected. As a result, the dashed line reaches values of $\condent(\rcn,\agen)$ close to $0.3$ [bit] in the leftmost part of the figure, while the uncertainty never falls below $0.4$ [bit] for $\radMax=125$ [m]. On the other hand, the availability of more transmitting devices leads to much more frequent deliveries. For the considered settings, with $\radMax=125$ [m], we get $\ps \simeq 0.1$, and a new reading is obtained on average every ${\sim} 9$ slots. This results in a distribution of $\condent(\rcn,\agen)$ which is much more concentrated around the value of uncertainty that follows a reset, and leads to a lower average conditional entropy (see \figr\ref{fig:HvsRad}).
Similar trends are observed for $\alpha=0.06$, with a shift towards higher uncertainty levels, due to the lower reliability of incoming messages.

\section{Conclusions}
\label{sec:conclusions}

This work provides an initial analysis of the uncertainty of an \ac{IoT} receiver over a spatio-temporal sensing task: along with the freshness dimension, which is traditionally accounted for through the \ac{AoI}, our analysis also includes a spatial component, assigning a lower reliability to updates from sensors farther from the center of a monitored area.
Our model of spatio-temporal freshness extends and complements results from the recent literature, which considered centralized models, by considering the optimization of a simple random access scheme, drawing some basic design guidelines.

Naturally, this is only a first step in the characterization of spatio-temporal freshness: the consideration of more complex estimators at the receiver, such as a hidden Markov model, would be an interesting avenue for further research, as would considering explicit information about each transmission's geolocation. Additionally, the study of different and more descriptive models for the source process and observation noise, such as the Ornstein-Uhlenbeck model, or even structural correlation models not purely tied to distance, would broaden the applicability of the results and provide further insights on the interplay of space, time, and value in remote monitoring.

\bibliographystyle{IEEEtran}
\bibliography{IEEEabrv,biblio_RandomAccess,biblio_AoI}

\begin{thebibliography}{10}
\providecommand{\url}[1]{#1}
\csname url@samestyle\endcsname
\providecommand{\newblock}{\relax}
\providecommand{\bibinfo}[2]{#2}
\providecommand{\BIBentrySTDinterwordspacing}{\spaceskip=0pt\relax}
\providecommand{\BIBentryALTinterwordstretchfactor}{4}
\providecommand{\BIBentryALTinterwordspacing}{\spaceskip=\fontdimen2\font plus
\BIBentryALTinterwordstretchfactor\fontdimen3\font minus
  \fontdimen4\font\relax}
\providecommand{\BIBforeignlanguage}[2]{{%
\expandafter\ifx\csname l@#1\endcsname\relax
\typeout{** WARNING: IEEEtran.bst: No hyphenation pattern has been}%
\typeout{** loaded for the language `#1'. Using the pattern for}%
\typeout{** the default language instead.}%
\else
\language=\csname l@#1\endcsname
\fi
#2}}
\providecommand{\BIBdecl}{\relax}
\BIBdecl

\bibitem{ericsson2024mobility}
Ericsson, ``Mobility report -- {November} 2024,'' Ericsson, Tech. Rep., 2024.

\bibitem{guo2021enabling}
F.~Guo, F.~R. Yu, H.~Zhang, X.~Li, H.~Ji, and V.~C. Leung, ``Enabling massive
  {IoT} toward {6G}: A comprehensive survey,'' \emph{IEEE Internet of Things
  J.}, vol.~8, no.~15, pp. 11\,891--11\,915, 2021.

\bibitem{Kaul11_SECON}
S.~{Kaul}, M.~{Gruteser}, V.~{Rai}, and J.~{Kenney}, ``Minimizing age of
  information in vehicular networks,'' in \emph{Proc. IEEE SECON}, June 2011.

\bibitem{Yates20_Survey}
R.~Yates, Y.~Sun, D.~Brown, S.~Kaul, E.~Modiano, and S.~Ulukus, ``Age of
  information: An introduction and survey,'' \emph{{IEEE} J. Sel. Areas
  Commun.}, vol.~39, no.~5, pp. 1183--1210, May 2021.

\bibitem{yang2021spatiotemporal}
H.~H. Yang, A.~Arafa, T.~Q. Quek, and H.~V. Poor, ``Spatiotemporal analysis for
  age of information in random access networks under last-come first-serve with
  replacement protocol,'' \emph{IEEE Trans. Wireless Commun.}, vol.~21, no.~4,
  pp. 2813--2829, 2021.

\bibitem{kalor2022timely}
A.~E. Kal{\o}r and P.~Popovski, ``Timely monitoring of dynamic sources with
  observations from multiple wireless sensors,'' \emph{IEEE/ACM Trans. Netw.},
  vol.~31, no.~3, pp. 1263--1276, 2022.

\bibitem{zancanaro2023modeling}
A.~Zancanaro, G.~Cisotto, and L.~Badia, ``Modeling value of information in
  remote sensing from correlated sources,'' \emph{Computer Communications},
  vol. 203, pp. 289--297, 2023.

\bibitem{tong2022age}
J.~Tong, L.~Fu, and Z.~Han, ``Age-of-information oriented scheduling for
  multichannel {IoT} systems with correlated sources,'' \emph{IEEE Trans.
  Wireless Commun.}, vol.~21, no.~11, pp. 9775--9790, 2022.

\bibitem{fidler20242d}
M.~Fidler, F.~Gallistl, J.~P. Champati, and J.~Widmer, ``{2D-AoI}:
  Age-of-information of distributed sensors for spatio-temporal processes,''
  \emph{arXiv preprint arXiv:2412.12789}, 2024.

\bibitem{bellavista2024odte}
P.~Bellavista, N.~Bicocchi, M.~Fogli, C.~Giannelli, M.~Mamei, and M.~Picone,
  ``{ODTE}: A metric for digital twin entanglement,'' \emph{IEEE Open J.
  Commun. Soc.}, 2024.

\bibitem{Abramson77:PacketBroadcasting}
N.~Abramson, ``The throughput of packet broadcasting channels,'' \emph{{IEEE}
  Trans. Commun.}, vol. COM-25, no.~1, pp. 117--128, 1977.

\bibitem{Liew22_TIT}
G.~Chen, S.-C. Liew, and Y.~Shao, ``Uncertainty-of-information scheduling: A
  restless multiarmed bandit framework,'' \emph{{IEEE} Trans. Inf. Theory},
  vol.~68, no.~9, pp. 6151--6173, 2022.

\bibitem{Yates17:AoI_SA}
R.~Yates and S.~Kaul, ``Status updates over unreliable multiaccess channels,''
  in \emph{Proc. IEEE ISIT}, 2017.

\bibitem{Munari21_TCOM_AoI}
A.~Munari, ``Modern random access: an age of information perspective on
  irregular repetition slotted {ALOHA},'' \emph{{IEEE} Trans. Commun.},
  vol.~69, no.~6, pp. 3572--3585, 2021.

\end{thebibliography}

\end{document}